\documentclass[10pt, final, twocolumn, oneside, conference]{IEEEtran}

\usepackage{graphicx}
\usepackage[printonlyused]{acronym}
\usepackage{xspace}
\usepackage{amsmath}
\usepackage{url}
\usepackage[dvipsnames]{xcolor}
\usepackage{multirow}
\usepackage{spreadtab}
\usepackage{color}
\usepackage{subfigure}
\usepackage{adjustbox}
\usepackage{listings}
\usepackage{color}
\usepackage{placeins}
\usepackage{adjustbox}
\usepackage{epstopdf}
\usepackage{float}
\usepackage{textcomp}
\usepackage{array}
\usepackage{color}
\usepackage{colortbl}
\usepackage{epstopdf}

\graphicspath{{./figures/}}

\begin{document}

\def \titolo {Smart Grids empowerment with Edge Computing: An Overview}
\title{\titolo}

\author{\IEEEauthorblockN{Pietro Boccadoro\IEEEauthorrefmark{1,2}}
	\IEEEauthorblockA{\IEEEauthorrefmark{1} Dep. of Electrical and Information Engineering (DEI), Politecnico di Bari, Bari, Italy\\
	\IEEEauthorrefmark{2}CNIT, Consorzio Nazionale Interuniversitario per le Telecomunicazioni, Bari, Italy}}

\maketitle

\acrodef{3GPP}{3rd Generation Partnership Project}
\acrodef{6TiSCH}{IPv6 over the TSCH mode of IEEE 802.15.4e}
\acrodef{AMI}{Advanced Monitoring System}
\acrodef{AMR}{Automatic Meter Reading}
\acrodef{BLE}{Bluetooth Low Energy}
\acrodef{BMS}{Building Management System}
\acrodef{DG}{Distributed Generator}
\acrodef{DSM}{Demand Side Management}
\acrodef{DSO}{Distribution System Operator}
\acrodef{EC}{Edge Computing}
\acrodef{GPRS}{General Packet Radio Service}
\acrodef{GPS}{Global Positioning System}
\acrodef{GSM}{Global System for Mobile}
\acrodef{HVAC}{Heating, Ventilation and Air Conditioning}
\acrodef{IoS}{Internet of Services}
\acrodef{IoT}{Internet of Things}
\acrodef{IPv6}{Internet Protocol version 6}
\acrodef{ISM}{Industrial, Scientific and Medical}
\acrodef{ITS}{Intelligent Transportation System}
\acrodef{LTE}{Long Term Evolution}
\acrodef{LoRa}{Long Range}
\acrodef{LPWAN}{Low Power Wide Area Network}
\acrodef{LV}{Low Voltage}
\acrodef{M2M}{Machine-to-Machine}
\acrodef{MV}{Medium Voltage}
\acrodef{NB-IoT}{Narrowband IoT}
\acrodef{NFC}{Near Field Communication}
\acrodef{PLC}{Power Line Communication}
\acrodef{PLR}{Packet Loss Ratio}
\acrodef{QoS}{Quality of Service}
\acrodef{RER}{Renewable Energy Resource}
\acrodef{RFID}{Radio Frequency IDentification}
\acrodef{SCADA}{Supervisory Control And Data Acquisition}
\acrodef{SDN}{Software Defined Network}
\acrodef{SG}{Smart Grid}

\begin{abstract}
Electric grids represent the angular stone of distribution networks. Since their introduction, a huge evolutionary process turned them from conventional electrical power network to advanced, real-time monitoring systems. In this process, the \ac{IoT} proved itself as a fast forwarding paradigm: smart devices, networks and communication protocol stacks are more and more integrated in a number of general purpose, industrial grade systems. In this framework, the upcoming \ac{IoS} is going to orchestrate the many sensors and components in \aclp{SG}, simultaneously enabling complex information management for energy suppliers, operators and consumers. The opportunity to continuously monitor and send important data (i.e., energy production, distribution, usage and storage) issues and facilitates the implementation of trailblazing functionalities. Of course, the more information are sent throughout the network, the higher the overload will be, thus resulting in a potentially worsening of the \ac{QoS} (i.e., communication latencies). This work overviews the concept of \aclp{SG} while investigating de-centralized computing and elaboration possibilities, leveraging the \ac{IoT} paradigm towards the \ac{IoS}.
\end{abstract}


\IEEEpeerreviewmaketitle

\section{Introduction}\label{introduction}
\aclp{SG} represent a technologically advanced ecosystem, able to carry out several tasks, spanning from continuous monitoring to \ac{DSM} \cite{SHAUKAT20181453}.
Such a colorful landscape is now accessible since \aclp{SG} have undergone a complex technological transformation process: in fact, since their first materialization as conventional systems, controlled by simple electro-mechanical devices, electric grids are now able to properly manage real-time monitoring tasks, \ac{M2M} communications, self calibration, maintenance scheduling, and pre-elaboration of data to be sent to remote servers \cite{SHAUKAT20181453}.
The employment of intelligent measurement and communication systems turned to a real need as it has been calculated that an important percentage (between 20 and 30\%) of the energy flowing throughout the whole conventional electric grid infrastructure is lost \cite{SHAUKAT20181453}. Due to substandard operations, the phenomenon may verify not only at generation side, but also during transmission and distribution.
The key factors involved in generation process are: (i) the cost of electricity, (ii) infrastructure aging, thus resulting in a less efficient distribution network, (iii) the carbon footprint, and (iv) green house gas emissions \cite{SHAUKAT20181453,7422663}.
Moreover, due to climate changes occurring over time, even the generation process is negatively affected.

To bridge the gap between the theoretical efficiency of such advanced systems and the actual one, many promising features characterize the desirable future of a \ac{SG}, such as: (a) intelligent de-centralized control, which implies (b) distributed intelligence, (c) resilience and fault-tolerance, (d) flexibility, (e) sustainability, (f) green energy employment, and (g) smart infrastructures \cite{7422663}.
The present contribution overviews the \aclp{SG} ecosystem, focusing on the main characteristics and problems \cite{8118364}. Afterward, it highlights how the advanced management functionalities \cite{8327976} can be specifically addressed relying on the \ac{IoT} paradigm.
Furthermore, the work addresses the \ac{EC} frontier (i.e., smart cities, industrial automation and smart metering), hereby considered as enabler for the upcoming \ac{IoS} scenario.

The present work is organized as follows: Section \ref{background} proposes an overview of \aclp{SG} and their peculiarities. Section \ref{communications} is dedicated to communication protocols and enabling paradigms that have been investigated as enablers. In Section \ref{challenges}, a detailed overview of challenges and open issues is given. Finally, Section \ref{conclusions} concludes the work and draws future work possibilities.

\section{Smart Grids in a nutshell}\label{background}
A typical \ac{SG} is able to carry out a number of tasks which require the improvement of both measurement and communication infrastructures of distribution networks \cite{SHAUKAT20181453}-\cite{QA17}.
Distributed sensing and measurement systems are needed to provide all necessary data for grid monitoring, control and management, as well as for the implementation of a number of smart functionalities, such as remote control of \aclp{DG}, real time analysis of power flows, \ac{AMR}, \ac{DSM}, and grid automation \cite{7422663}.
Such renovation has direct effects on power lines communications for \ac{SG} monitoring and management in both \ac{MV} and \ac{LV} distribution \cite{8327976}. 
The typical electrical network variables that are taken into account in monitoring activities are (i) powers, (ii) voltages, and (iii) currents.
Moreover, advanced monitoring may include environmental parameters (i.e., temperature), \aclp{DG} power production, remote commands and status control signals (such as security or safety warning/threshold levels) \cite{8118364}.
The generation process results to be sensibly affected by: (i) the cost of electricity, (ii) the aging of the infrastructure, thus resulting in a less efficient distribution network, (iii) the carbon footprint, and (iv) green house gas emissions \cite{SHAUKAT20181453,7422663}.
These data are exchanged among the different entities of a \ac{SG} structure, such as distribution operators, active users, and prosumers (for instance, users that are both accessing data and producing their own).
\aclp{SG} functional requirements are conceived as referred to both capillary distribution network working conditions (e.g., overloads) \cite{7422663,7995239}.
One of the solutions that are frequently adopted is the employment of wireless technologies, \ac{GSM} or \ac{PLC} systems.
The latter possibility proposes low installation and operational costs, intrinsic security from cyber-attacks, and direct and complete control by \aclp{DSO}. In fact, power lines are already located in many territories and are owned by the \aclp{DSO}, thus communication provider costs are avoided and potential intruders would encounter difficulties in accessing the network.

Among the promising features that make \ac{SG} reliable, cost-effective and easy to be installed, there are: (a) intelligent de-centralized control, (b) resilience and fault-tolerance, (c) flexibility, (d) sustainablity, (e) digitalized, (f) distributed intelligence, (g) green energy employment, and (h) smart infrastructure \cite{SHAUKAT20181453}.
\begin{figure}[b]
	\centering
	\includegraphics[width=0.5\textwidth]{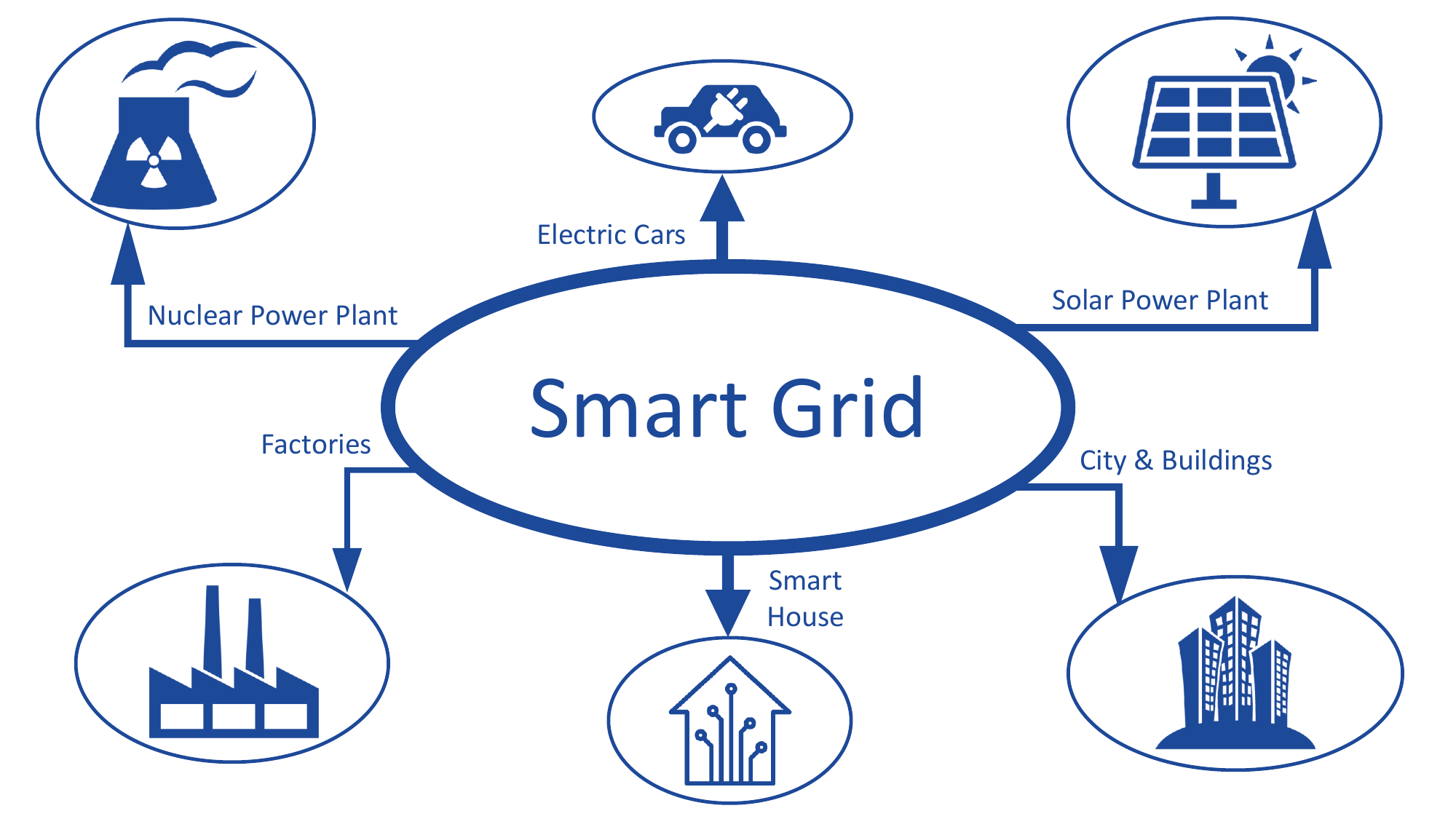}
	\caption{Smart Grids integration perspectives.}
	\label{fig:smart_grid}
\end{figure}
On the consumer's side, functionalities empowerment and continuous monitoring activities can be carried out in a closed-loop fashion, thus granting user experience while maximizing the optimization results. Moreover, consumer empowerment specifically addresses \ac{DSM} with the long term objective of lowering energy footprint. This result can be achieved since the customer becomes increasingly aware of the way electricity is used and gains a central role in stating whether \aclp{BMS} are working properly.
In Figure \ref{fig:smart_grid}, \aclp{SG}' employability is represented in various application fields.

\section{IoT protocols for Smart Grids}\label{communications}
The \ac{IoT} is a broad concept that can be applied to the whole communication network, starting from the end-devices (typically small-sized, embedded systems with several sensors on-board and limited computational and storage capabilities) to application services exposed through web interfaces for users willing to perform data analysis and representation.
This is so as \ac{IoT} devices embed central processing units, elaboration and storage memories, peripheral devices (i.e., sensors and/or actuators), and Radio-Frequency transceiver, thus efficiently allowing \ac{M2M} communication \cite{8327976,7815384}.
One of the possible rationale to distinguish and categorize \ac{IoT} protocols stacks is the communications range, which spans from few centimeters (e.g., \ac{RFID}, \ac{NFC}) to tens of kilometers in \ac{LPWAN} scenarios (e.g., \ac{LTE}, \ac{LoRa}, \ac{3GPP} \ac{NB-IoT} and Sigfox). Short range communications can be enabled by standardized solutions, for instance \ac{BLE} or IEEE 802.15.4 \cite{7815384,SONG2017460}.
\aclp{LPWAN} can overcome the barrier of physical position (generally indicated through \ac{GPS} coordinates) and system sophistication, keeping the complexity of the systems as low as possible for the end-users. For example, \ac{NB-IoT} is inherited from cellular communication, and seamlessly works over those technologies (i.e., \ac{GSM} and \ac{LTE}) in licensed frequency bands. \ac{LoRa} technology, instead, operates in the unlicensed frequency band, thus proposing less limitations and a lower time-to-market perspective. Therefore, \ac{LoRa} technology is perfect for outlying regions without cellular network coverage, or for establishing private networks with specific requirements for quality and security.
\begin{table*}[h]
	\centering
	\begin{tabular}{l|c|c|c|}
		\cline{2-4}
		& \textbf{LoRaWAN}                                                                                         & \textbf{Sigfox}                                                                                          & \textbf{NB-IoT}                                                                                                                      \\ \hline
		\multicolumn{1}{|l|}{\textbf{Governing Body}}       & LoRa Alliance                                                                                            & Sigfox                                                                                                   & Standardized                                                                                                                         \\ \hline
		\multicolumn{1}{|l|}{\textbf{Frequency}}            & \begin{tabular}[c]{@{}c@{}}868 MHz (Europe)\\ 915 MHz (North America)\\ Unlicenced spectrum\end{tabular} & \begin{tabular}[c]{@{}c@{}}868 MHz (Europe)\\ 915 MHz (North America)\\ Unlicenced spectrum\end{tabular} & \begin{tabular}[c]{@{}c@{}}5 different bands - 5 HW modules,\\ 3 different European bands\\ Licenced spectrum\end{tabular} \\ \hline
		\multicolumn{1}{|l|}{\textbf{Bandwidth}}            & 125 kHz                                                                                                  & 100 Hz                                                                                                   & 180 kHz                                                                                                                              \\ \hline
		\multicolumn{1}{|l|}{\textbf{Range}}                & \begin{tabular}[c]{@{}c@{}}2 - 5 km (urban)\\ 15 km (rural)\end{tabular}                                 & \begin{tabular}[c]{@{}c@{}}3 - 10 km (urban)\\ 30 - 50 km (rural)\end{tabular}                           & 15 km                                                                                                                                \\ \hline
		\multicolumn{1}{|l|}{\textbf{Coverage (dB)}}        & 157 dB                                                                                                   & 149 dB                                                                                                   & 164 dB                                                                                                                               \\ \hline
		\multicolumn{1}{|l|}{\textbf{Uplink Bandwidth}}     & \begin{tabular}[c]{@{}c@{}}EU: 0.3 - 50 kbps\\ US: 500x10 bytes/day\end{tabular}                         & \begin{tabular}[c]{@{}c@{}}100 bps\\ 140x12 butes/day\end{tabular}                                       & 250 kbps                                                                                                                             \\ \hline
		\multicolumn{1}{|l|}{\textbf{Downlink Bandwidth}}   & \begin{tabular}[c]{@{}c@{}}EU: 0.3 - 50 kbps\\ US: 500x10 bytes/day\end{tabular}                         & 4x8 bytes/day                                                                                            & 20 kbps                                                                                                                              \\ \hline
		\multicolumn{1}{|l|}{\textbf{Bidirectional}}        & Class dependent                                                                                          & No                                                                                                       & Yes                                                                                                                                  \\ \hline
		\multicolumn{1}{|l|}{\textbf{Device per AP}}        & \begin{tabular}[c]{@{}c@{}}UL: \textgreater 1 M\\ DL: \textgreater 100 k\end{tabular}                    & \textless  1 M                                                                                           & 50k                                                                                                                                  \\ \hline
		\multicolumn{1}{|l|}{\textbf{Packet size}}          & Defined by user                                                                                          & 12 bytes                                                                                                 &                                                                                                                                      \\ \hline
		\multicolumn{1}{|l|}{\textbf{Battery Lifetime}}     & +10 years                                                                                                & +10 years                                                                                                & +10 years                                                                                                                            \\ \hline
		\multicolumn{1}{|l|}{\textbf{Security}}             & 32 bit, multilayer AES-128                                                                               & 16 bit                                                                                                   & \ac{3GPP} (128-256 bit)                                                                                                                    \\ \hline
		\multicolumn{1}{|l|}{\textbf{Localization support}} & Yes                                                                                                      & No                                                                                                       & Needs \ac{GPS}                                                                                                                            \\ \hline
		\multicolumn{1}{|l|}{\textbf{Applications}} & \begin{tabular}[c]{@{}c@{}}Smart Agriculture\\ Smart Cities\\ Utilities\end{tabular}    & \begin{tabular}[c]{@{}c@{}}Smart parking \& traffic\\ Smart Environment\\ M2M\end{tabular}               & \begin{tabular}[c]{@{}c@{}}Industrial plant monitoring\\ Smart metering\\ eHealth\end{tabular}                                 \\ \hline
	\end{tabular}
	\caption{Technical specification of the main \aclp{LPWAN} protocol stacks.}
	\label{lpwan_techs}
\end{table*}

The primary advantages of employing \ac{LPWAN} solutions lie in the possibilities enabled by the wide communication coverage area and the jointly granted low energy footprint. 
Pragmatically speaking, when all the production and distribution systems are online, energy flowing through the network, as well as heat and gas in \ac{HVAC} systems, may create harsh operative conditions, which impairs sensing and communications.
The low-power mode of the involved \ac{IoT} embedded systems grant long-term reliability and their reachability in wide-spread topologies enables highly efficient data collections and inter-operations.
On the other end, typical \ac{LPWAN} protocol stacks only account for low data rate. Anyway, such a downside is anyhow limited by the typical \ac{IoT} constraints in terms of computational and storage capabilities. Moreover, the problem can be efficiently solved optimizing communication keeping the amount of data to the protocol upper-bound and scheduling data transmissions and receptions using duty-cycling policies.
Table \ref{lpwan_techs} summarizes the main characteristics of some \ac{LPWAN} solutions.

The aforementioned communication technologies may be successfully applied to the \ac{SG} ecosystem as the production, distribution and usage chain can be functionally analyzed. In particular, on the demand side, communication infrastructure can be significantly ameliorated at the power distribution level, lowering the amount of devices and unifying technological solutions. The same happens for utilization systems, at \ac{LV} levels.
Several challenging aspects are also proposed as to solve the issue of \ac{DSM}; in particular, the power automation architectures investigated so far have been design properly addressing needs, standards and requirements of centralized generation and transmission systems. Nevertheless, those indications are increasingly difficult to be satisfied as the network integrates distributed energy resources.

\section{Challenges and open issues}\label{challenges}
The \ac{IoS} is conceived as a characterizing information paradigm and may successfully qualify the evolution of \aclp{SG}. Leveraging the potential of \ac{IoT} design, software applications, development and delivery platforms, the complexity and infrastructural details could still be kept hidden for end-users \cite{6203493}.
In this scenario, cloud technologies \cite{SONG2017460} offer interesting abstraction possibilities, while comprising different provisioning models for on-demand access, such as:
\begin{itemize}
	\item service elasticity, granted by the seamless accessibility and resulting from cloud's scalability of envisioned services;
	\item infrastructure and platforms cost reduction, resulting from the adaptation to service demands;
	\item pay-per-use models, from which the time-to-market may results to be significantly improved;
	\item increased service availability and reliability, resulting from the replication of service components and rapid deployment of new service instances.
\end{itemize}
It is worth noting that a cloud infrastructure is interoperable by design, and enables users to deploy even a single service on multiple clouds instances.
\aclp{SG} do not only apply to industrial field. Indeed, the different application domains involved are three-folded:

\begin{itemize}
	\item \emph{Customer Domain}.
	On customer side, users wants \aclp{SG} not to operate on a flat rate, but carring out dynamic adjustment operations to control the energy utilization (and utilization costs).
	The monitoring of the values involved in such activities are part of a bi-direction flow of information that occurs between utilities and prosumers \cite{7995239}.
	Customers may benefit from the upcoming \ac{IoS} in terms of smart appliances and smart homes, where the main goal is to achieve a reduction in energy consumption and bills.	
	\item \emph{Information and Communication Domain}.
	The profitability of the market of smart energy supply strongly depends on appropriate communication channels and infrastructures. A possible enhancement in this field is the substitution of the current IEEE 802.15.4 (i.e., ZigBee, or \ac{6TiSCH}) or Wi-Fi based solution in favour of \aclp{LPWAN} ones, which natively support and integrate a wireless-to-cloud solution.\\
	The active distribution networks are currently experiencing the need for a capillary generation and storage at \ac{LV} level. Several smart energy devices now request the level of information-sharing that is associated with distribution system operation and an aggregated demand response.
	Nevertheless, it is difficult to extend current information systems to end consumers due to complex field environments and security concerns.
	Utilization networks at the lowest voltage level lack the space and communication channels required to add additional remote devices to enable interoperations with power distribution systems.\\
	Another recent frontier in this domain is represented by electric vehicles towards aggregated vehicle-to-grid operations. Even if electric vehicles are currently accepted as the transportation solution of the future, their large-scale integration is still forbidden. The limiting factors are related to inherent operational flexibility of energy systems. The road ahead to a completely integrated vehicle-to-grid framework could be properly sustained by power systems balancing control. Frequency regulation and peak shaving are among the main functionalities that smart metering could release.	
	\item \emph{Grid Domain}.
	Micro-Grids and distributed energy systems are developed with the clear aim of increasing the percentage of clean and sustainable energy sources, while trying to grant the power supply to be as stable and reliable as possible. The applicability of such paradigm is mainly referred to energy districts, spanning from a river valley to a whole island \cite{8118364}.
	The multi-energy system deriving from the integration of renewable energy sources clearly leads to the need to design several energy conversion plants (e.g. power-to-gas, power-to-heat, and combined heat and power systems). The most promising way to reach this goal of improving energy utilization efficiency is the integration into energy hubs.
\end{itemize}

\section{\acl{EC}: a key enabler}
In the context of \aclp{SG}, several possibilities are now forbidden. First of all, the de-centralized architecture of the monitoring system lacks in terms of self-awareness and self-organization. Even if they account for a large quota of the advanced sensing and elaboration tasks, \aclp{SG} are not aware of network topology. Furthermore, dynamic optimization still remains one of the most interesting research challenges. Among them, machine learning algorithms can be employed in order to perform condition-base monitoring and carry out prognostic activities (e.g., predictive maintenance tasks), in order to evolve the less efficient run-to-failure or preventive maintenance approaches.

The \ac{IoS} perspective strongly addresses communication technologies, such as wireless \ac{AMI}, Phasor Measurement Unit (PMU), \ac{SCADA}, and \ac{M2M} communications. 
Moreover, energy control in the \ac{EC} refers to architectural model focusing on consumer empowerment, Demand Response Program (DRP), and \ac{DSM} \cite{6203493}.
The cloud computing itself is not sufficient to meet \ac{QoS} requirements such as low latency, location awareness, and mobility support. In this context, \ac{EC} was introduced to bring both cloud services and resources closer to the user.
Being a reference architectural paradigm, \ac{EC} candidates itself as key enabler for *-Grid plants, spanning from Pico to Micro. Moreover, it may properly serve Inter Grids, Virtual Power Plants, and \aclp{DG}.

\begin{figure}[htbp]
	\centering
	\includegraphics[width=\columnwidth]{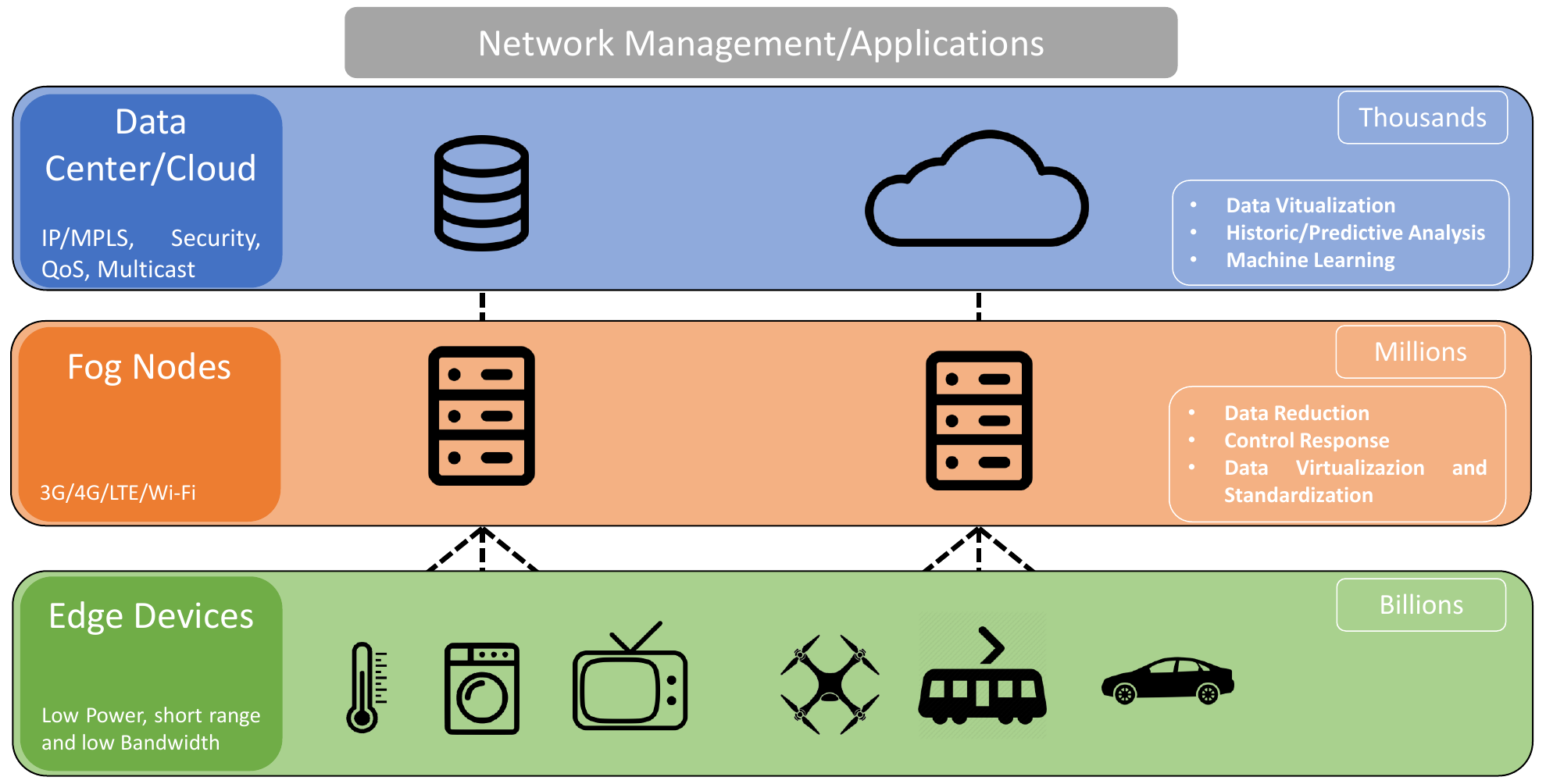}
	\caption{High-level description of distributed \ac{EC} architecture.}
	\label{fig:edge}
\end{figure}

A research perspective in the \ac{EC} scenario is represented by the integration of \aclp{RER} within the current \aclp{SG}. In particular, efforts could be devoted in solving the issues related to \ac{DG}, consumer empowerment and prosumers interaction \cite{SONG2017460}.
Among the many, \ac{EC} tackles entities (such as, smart meters, \ac{IoT} devices, cloud networking, Datacenters and identity) federation and architectural challenges. The latter envision \aclp{SDN} and \ac{IPv6}-based monitoring systems working together to create an homogeneous infrastructure, starting from the fragmented one that exists today.
Functionally speaking, \ac{EC} on a distributed architecture will strongly address scheduling and fault tolerance challenges.
In figure \ref{fig:edge}, the high-level interactions are depicted. In particular, it is proposed a vision of edge devices, mainly performing data gathering, communicating with fog nodes. Once data have been communicated, pre-elaboration and virtualization tasks are carried out in order to facilitate data center and cloud instances/services in exposing structured information.

\section{Conclusions}\label{conclusions}
This work addresses \ac{SG} and their main characteristics, which have been investigated in terms of both monitoring functionalities and communication capabilities.
The new and heterogeneous features of \aclp{SG} strongly deserve additional effort to get to the point of de-centralized and distributed intelligence.
To the undergoing evolutionary process of \aclp{SG}, the \ac{IoT} contributes enhancing communications. Among the many \ac{LPWAN} possibilities, \ac{NB-IoT} and \ac{LoRa} technologies can be considered as breakthroughs in their field. In fact, compared with cellular networks, the two account for (i) a lower quota of energy consumed, (ii) improved coverage area, (iii) optimized transmission and reception operations, and (iv) the concrete possibility to distribute elaboration tasks on the borders of the network.
Moving on to the end of the paper, several open issues and technological challenges have been spotted out, thus envisioning future research possibilities in both academic and industrial contexts.
\ac{EC} is not surprisingly pushing the \ac{SG} to a brand new level. Selective offloading scheme can satisfy the latency requirements of different services and reduce the energy consumption of IoT device, even for the most efficient.
Future researches will go deeper into security aspects of \ac{SG}. In fact, cryptography and security algorithms can be successfully integrated to perform secure remote monitoring \cite{BD14}. The long-term perspective of such research will pave the way to a better environment for future generations, thanks to reliable \aclp{SG} which will be able to lower the costs for customers, carbon-dioxide emissions, and global warming.


\bibliographystyle{IEEEtran}
\bibliography{references}

\end{document}